\documentclass[aps,prd,reprint,superscriptaddress,nofootinbib]{revtex4-2}

\usepackage{amsmath,amssymb,bm}
\usepackage{graphicx}
\usepackage{booktabs}
\usepackage{array}
\usepackage[colorlinks=true,allcolors=blue]{hyperref}

\begin{document}

\title{Equilibrium Gibbs Bifurcations of Bardeen-AdS Black Holes at Fixed Pressure}
\author{J.-K. Wang}
\email{WangTheoPhys@outlook.com}
\affiliation{Tsinghua University, Beijing 100084, China}
\affiliation{TsingyuAI Corporation, Beijing 100084, China}
\date{May 22, 2026}

\begin{abstract}
\begin{center}
AI Agent: \textbf{Scientify} by \textbf{TsingyuAI} (\href{https://scientify.tech}{Scientify.tech})
\end{center}
In the direct horizon convention, the fixed-pressure Gibbs curve of the four-dimensional Bardeen-AdS black hole passes through an intermediate sequence between the Reissner-Nordstrom-AdS swallow-tail class and the single-branch regime as the regularization scale is increased. The on-shell curve is classified by its turning points and self-intersections, followed by local heat-capacity filtering and construction of the lower Gibbs envelope over stable branches. The deformation is resolved into three boundaries: $g_*(P)$, where the RN-AdS-like topology is lost; $g_c(P)$, where the c-shaped sector begins; and $g_s(P)$, where the positive-temperature multibranch structure terminates. A reduced-variable analysis shows that these boundaries are controlled by the dimensionless combination $8\pi P g^2$, accounting for their inverse-square-root pressure dependence and giving an analytic value for the final single-branch boundary. The equilibrium construction further shows that stable small/large coexistence can survive the first topology change, whereas the representative c-shaped regime has no stable crossing. Within the direct convention, these results define a Gibbs bifurcation structure for Bardeen-AdS thermodynamics.

\end{abstract}

\maketitle
\begingroup
\renewcommand{\thefootnote}{}
\footnotetext{Calculations, numerical analysis, and manuscript preparation were performed by Scientify, a continuous-knowledge-metabolism AI assistant, working under human scientific supervision.}
\endgroup

\section{Introduction}
The thermodynamics of asymptotically anti-de Sitter black holes relates horizon data to global phase structure through the Gibbs free energy. In the four-dimensional RN-AdS solution, the fixed-pressure Gibbs curve develops the standard swallow-tail below the critical pressure, giving the small/large black-hole transition \cite{hawkingpage1983,chamblin1999a,chamblin1999b,kubiznak2012,mann2017}. Here RN-AdS serves as a reference black-hole solution with a well-defined one-intersection Gibbs topology. The direct Bardeen family considered below has a different limiting geometry.

Regular black holes introduce a related issue: singularity resolution can alter the global organization of thermodynamic branches. The Bardeen-AdS black hole is the standard prototype of a regular black-hole geometry \cite{bardeen1968,ayonbeato2000}; it admits explicit horizon thermodynamics and contains a regularization parameter $g$ that affects the small-radius sector where additional branches can occur \cite{li2019,tzikas2019,guomiao2022}.

Regular-black-hole thermodynamics is also prescription dependent. Different choices of entropy, volume, and thermodynamic variables can give inequivalent phase spaces \cite{ma2014,guomiao2022,simovic2024}. It has been shown that the standard van der Waals interpretation can fail for regular AdS black holes, including in the use of the equal-area law and in the identification of critical points \cite{fan2017}. This prescription dependence appears in recent analyses: constrained singular-to-regular constructions can replace the standard swallow-tail by $8$-shaped or c-shaped Gibbs curves \cite{ma2025}, while restricted phase-space thermodynamics can keep Bardeen phase behavior close to the RN-AdS pattern \cite{ladghami2025}. Regularization has also been formulated inside a broader theory family, avoiding the imposition of an RN-type limit on every regular solution \cite{hennigar2025}.

The thermodynamic convention is fixed before comparing Gibbs topologies. The variables used below are the metric enthalpy, the surface-gravity temperature, the area-law entropy, and the Gibbs potential $G=M-TS$. At fixed pressure, the relevant values of $g$ are those at which the Bardeen-AdS Gibbs curve leaves the RN-AdS-like one-intersection swallow-tail topology. After the rescaling used below, the fixed-pressure problem becomes a one-parameter bifurcation problem.

The analysis separates the topology of the parametric Gibbs projection from the equilibrium phase structure obtained after applying the heat-capacity criterion and taking the lower Gibbs envelope over locally stable branches. The raw topology is classified as RN-AdS-like, $8$-shaped, c-shaped, or single-branch. A change in this projection can occur while stable coexistence remains present.

The reduced variables give analytic control of the boundary scaling. After rescaling $r_+=g\rho$, the direct Bardeen-AdS temperature, the direct Gibbs potential, and the turning-point equation all depend only on the combination
\begin{equation}
\lambda = 8\pi P g^2.
\end{equation}
This scaling accounts for the inverse-square-root form of the boundary curves in the $P$-$g$ plane. The exponent follows from dimensional reduction, while the constants $\lambda_*$, $\lambda_c$, and $\lambda_s$ determine the bifurcation sequence.

The paper is organized as follows. Section~II defines the thermodynamic setup and the RN-AdS reference black-hole solution. Sections~III--VI give the Bardeen-AdS Gibbs construction, the local-stability and equilibrium-envelope analysis, and the reduced-variable form of the turning-point problem. Section~VII presents the numerical thresholds, followed by the conclusion.

\section{Thermodynamic setup}
For the four-dimensional RN-AdS reference black hole we fix $Q=1$ and identify the pressure as
\begin{equation}
P=\frac{3}{8\pi l^2}.
\end{equation}
The enthalpy, Hawking temperature, entropy, and Gibbs free energy are
\begin{align}
M_{\rm RN} &= \frac{r_+}{2}+\frac{Q^2}{2r_+}+\frac{4\pi P r_+^3}{3}, \\
T_{\rm RN} &= \frac{1+8\pi P r_+^2-Q^2/r_+^2}{4\pi r_+}, \\
S_{\rm RN} &= \pi r_+^2, \\
G_{\rm RN} &= M_{\rm RN}-T_{\rm RN}S_{\rm RN}.
\end{align}
The critical pressure is $P_c=1/(96\pi Q^2)$, and in the reference figure we take $P/P_c=0.6$.

For the direct Bardeen-AdS black hole we use
\begin{equation}
f(r)=1-\frac{2Mr^2}{(r^2+g^2)^{3/2}}+\frac{8\pi P r^2}{3},
\end{equation}
where $g$ is the regularization parameter. Solving the horizon condition $f(r_+)=0$ for $M$ yields
\begin{equation}
M_{\rm B}=\frac{(r_+^2+g^2)^{3/2}(3+8\pi P r_+^2)}{6r_+^2},
\end{equation}
while the surface-gravity temperature is
\begin{equation}
T_{\rm B}=\frac{r_+^2-2g^2+8\pi P r_+^4}{4\pi r_+(r_+^2+g^2)}.
\end{equation}
We retain the area law,
\begin{equation}
S_{\rm B}=\pi r_+^2,
\end{equation}
and define the direct Gibbs potential by
\begin{equation}
G_{\rm B}=M_{\rm B}-T_{\rm B}S_{\rm B}.
\end{equation}

These definitions specify the direct horizon-based thermodynamic convention in which the fixed-pressure threshold problem is posed.

\section{RN-AdS reference black hole}
The RN-AdS reference solution fixes the topology used for comparison. In the fixed-pressure ensemble, its temperature
\begin{equation}
T_{\rm RN}(r_+)=\frac{1}{4\pi r_+}\left(1-\frac{Q^2}{r_+^2}+8\pi P r_+^2\right)
\end{equation}
can possess two positive turning points. These turning points correspond to three black-hole branches, and the corresponding Gibbs projection has the standard one-intersection swallow-tail.

Introducing $x=r_+^2$, the turning-point condition becomes
\begin{equation}
8\pi P x^2-x+3Q^2=0
\end{equation}
which follows from $dT_{\rm RN}/dr_+=0$.
This quadratic encodes the loss of the RN-AdS branch structure above the critical pressure. The Bardeen deformation changes the turning-point polynomial, allowing the topology of the Gibbs projection to change at fixed pressure.

Accordingly, the term ``RN-AdS-like'' is used in a topological sense: the Bardeen-AdS Gibbs curve lies in the same one-intersection swallow-tail class as the RN-AdS reference black hole. The statement concerns the Gibbs projection rather than a $g\to0$ metric limit.

\section{Direct Bardeen-AdS thermodynamics}
The Bardeen-AdS temperature in the direct convention is
\begin{equation}
T_{\rm B}(r_+;P,g)=\frac{8\pi P r_+^4+r_+^2-2g^2}{4\pi r_+(r_+^2+g^2)}.
\end{equation}
Compared with the RN-AdS reference black hole, the regularization scale enters both the numerator and the denominator. It changes the branch structure as well as the critical scale.

The direct Gibbs potential in this convention is
\begin{equation}
G_{\rm B}(r_+;P,g)=M_{\rm B}-T_{\rm B}(r_+;P,g)\,\pi r_+^2.
\end{equation}
In this form, the Gibbs curve is the parametric curve
\begin{equation}
\bigl(T_{\rm B}(r_+;P,g),\,G_{\rm B}(r_+;P,g)\bigr),
\end{equation}
with $r_+$ restricted to the positive-temperature domain. The fixed-pressure thermodynamic problem is then determined by the branch structure of this curve.

The turning points of $T(r_+)$ determine the monotonic temperature branches and provide the starting data for the Gibbs-curve classification.

\section{Local stability and equilibrium Gibbs envelope}
For the direct Bardeen-AdS black hole in extended phase space, with $(P,g)$ fixed, local thermodynamic stability is governed by the heat capacity at fixed pressure,
\begin{equation}
C_P \equiv T\left(\frac{\partial S}{\partial T}\right)_{P,g}.
\end{equation}
Using the area-law entropy $S=\pi r_+^2$ and the direct temperature $T_{\rm B}(r_+;P,g)$, one finds
\begin{equation}
C_P=
\frac{2\pi r_+^{2}(r_+^{2}+g^{2})\left(8\pi P r_+^{4}+r_+^{2}-2g^{2}\right)}
{8\pi P r_+^{6}+(24\pi P g^{2}-1)r_+^{4}+7g^{2}r_+^{2}+2g^{4}}.
\label{eq:cp-main}
\end{equation}
The sign of $C_P$ distinguishes locally stable and unstable branches: $C_P>0$ indicates local stability, while $C_P<0$ indicates instability.

The turning points of $T(r_+)$ are spinodals of the direct thermodynamic system, since
\begin{equation}
C_P^{-1}=0
\end{equation}
is the fixed-pressure stability condition at $\left(\partial T/\partial r_+\right)_{P,g}=0$. Thus the turning-point structure is tied directly to local thermodynamic stability.

At fixed $(T,P,g)$, the globally preferred equilibrium configuration is determined by the lower envelope of the stable branches. We define
\begin{equation}
G_{\rm eq}(T;P,g)=\min_i G_i(T;P,g),
\end{equation}
where the minimum is taken only over branches with $C_P>0$. In the implementation, the positive-temperature branches are filtered by the stability condition $C_P>0$, interpolated onto a common temperature grid, and minimized pointwise in Gibbs free energy.

Stable coexistence between small and large black-hole branches is located by the usual conditions
\begin{equation}
T(r_s;P,g)=T(r_l;P,g),
\qquad
G(r_s;P,g)=G(r_l;P,g),
\end{equation}
with latent heat
\begin{equation}
L=T\,[S(r_l)-S(r_s)].
\end{equation}
These conditions define the equilibrium coexistence problem independently of the visual topology of the raw parametric curve.

Stable branch competition can persist beyond the first topology boundary at $g_*(P)$. The lower-envelope construction therefore separates local Gibbs-curve topology from equilibrium coexistence.

\section{Reduced variables and bifurcation structure}
The reduced variables
\begin{equation}
r_+=g\rho,
\qquad
\lambda=8\pi P g^2.
\end{equation}
put the direct temperature in the form
\begin{equation}
gT_{\rm B}=\frac{\rho^2-2+\lambda \rho^4}{4\pi \rho(\rho^2+1)},
\end{equation}
and the reduced Gibbs potential is
\begin{equation}
\frac{G_{\rm B}}{g}
=\frac{(\rho^2+1)^{3/2}(3+\lambda \rho^2)}{6\rho^2}
-\frac{\rho(\rho^2-2+\lambda \rho^4)}{4(\rho^2+1)}.
\end{equation}
The reduced $G$-$T$ curve therefore depends on $P$ and $g$ only through $\lambda$.

Thus an inverse-square-root boundary in the $P$-$g$ plane follows from the reduced variables. If a topology boundary occurs at $\lambda=\lambda_i$, then
\begin{equation}
g_i(P)=\sqrt{\frac{\lambda_i}{8\pi}}\,P^{-1/2}.
\end{equation}
The exponent is fixed by the reduced variables. The constants $\lambda_*$, $\lambda_c$, and $\lambda_s$ specify the bifurcation sequence.

Turning points are determined by the positive roots of $dT/dr_+=0$. In the original variable $x=r_+^2$ this gives
\begin{equation}
8\pi P x^3 +(24\pi P g^2-1)x^2 +7g^2 x +2g^4 =0,
\end{equation}
while in the reduced variable $y=x/g^2$ it becomes
\begin{equation}
\lambda y^3 +(3\lambda-1)y^2 +7y +2 =0.
\label{eq:reduced-cubic}
\end{equation}
The turning-point structure is therefore controlled by the same reduced parameter as the Gibbs curve.

The single-branch boundary follows from the discriminant of Eq.~(\ref{eq:reduced-cubic}). Writing
\begin{equation}
\Delta_y=\mathrm{Disc}_y\!\left[\lambda y^3 +(3\lambda-1)y^2 +7y +2\right],
\end{equation}
one finds
\begin{equation}
\Delta_y=-(\lambda-3)\bigl(216\lambda^2-657\lambda+19\bigr).
\end{equation}
The physically relevant turning-point merger occurs at the smaller positive root
\begin{equation}
\lambda_s=\frac{73}{48}-\frac{13\sqrt{273}}{144},
\end{equation}
with numerical value $\lambda_s=0.02919964344$. The corresponding boundary is
\begin{equation}
g_s(P)=\sqrt{\frac{\lambda_s}{8\pi P}}
=0.03408543527\,P^{-1/2},
\end{equation}
which gives the analytic single-branch boundary in the direct convention.

The remaining boundaries, $\lambda_*$ and $\lambda_c$, are obtained by continuation of the topology labels and then checked through the reduced-variable collapse. For the exact pressure slices,
\begin{equation}
\lambda_* = 6.4116\times 10^{-5},
\qquad
\lambda_c = 1.97807\times 10^{-2},
\end{equation}
with variations below the last retained digit across the pressure slices. The ordered sequence
\begin{equation}
\lambda_* < \lambda_c < \lambda_s
\end{equation}
separates the direct Bardeen-AdS problem into three regimes: loss of RN-AdS-like topology, transition out of the $8$-shaped sector into the c-shaped sector, and final entry into the single-branch regime.

\section{Same-family singular versus regular interpretation}
In the direct Bardeen-AdS construction adopted here, varying the regularization parameter $g$ changes the regular core within a family whose $g\to0$ limit is Schwarzschild-AdS. The comparison with RN-AdS is therefore thermodynamic and topological.

With this convention fixed, RN-AdS provides a charged-AdS reference for comparing the organization of turning points, coexistence structure, and Gibbs morphology. The parent family of the direct Bardeen-AdS geometry is different, so the direct solution is treated on its own thermodynamic footing.

This interpretation is consistent with the broader literature. Charged black-hole families can realize regularity on constrained slices of larger parameter spaces \cite{fanwang2016}. A corrected first law can also be required when the mass parameter enters the matter sector nontrivially \cite{ma2014}. In singular-to-regular constructions based on a singular ``mother'' solution, the regular Bardeen-AdS black hole can exhibit $8$-shaped and c-shaped Gibbs structures in place of the standard swallow-tail \cite{ma2025}. At the same time, restricted phase-space thermodynamics of Bardeen black holes can retain phase behavior close to RN-AdS \cite{ladghami2025}. Together with broader analyses of regular AdS black holes \cite{hennigar2025}, these results place singularity resolution and thermodynamic organization inside a larger space of prescriptions and theory families.

For the direct horizon-based Bardeen-AdS convention, the fixed-pressure Gibbs projection undergoes a reproducible sequence of equilibrium-relevant bifurcations controlled by a single reduced parameter $\lambda=8\pi P g^2$.

\section{Numerical results}
\subsection{RN-AdS reference black hole}

\begin{figure}[t]
\centering
\includegraphics[width=0.95\linewidth]{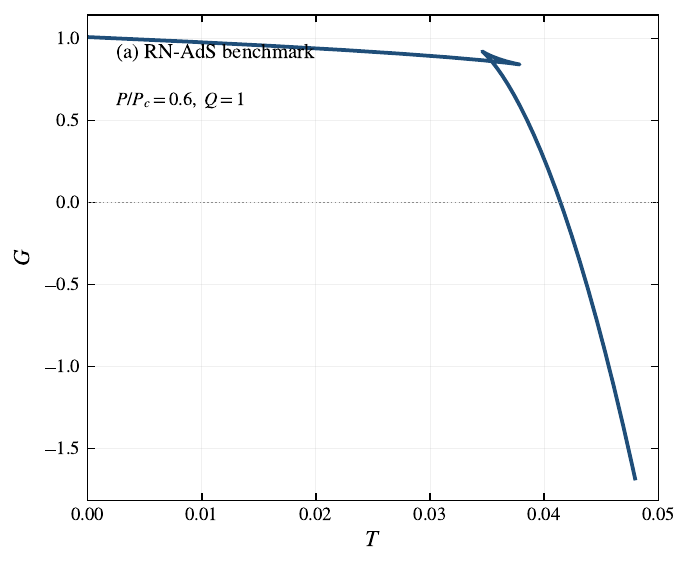}
\caption{RN-AdS reference black hole at $P/P_c=0.6$ with fixed $Q=1$. The Gibbs curve exhibits the standard one-intersection swallow-tail.}
\label{fig:rn-}
\end{figure}

The RN-AdS reference curve in Fig.~\ref{fig:rn-} reproduces the standard one-intersection swallow-tail. Within the same classifier used later for Bardeen-AdS, the curve has two positive turning points and one branch intersection. This reference defines the operational meaning of ``RN-AdS-like'' and sets the comparison class for the direct Bardeen-AdS curve.

\subsection{Representative raw topology changes}

\begin{figure*}[t]
\centering
\includegraphics[width=0.98\textwidth]{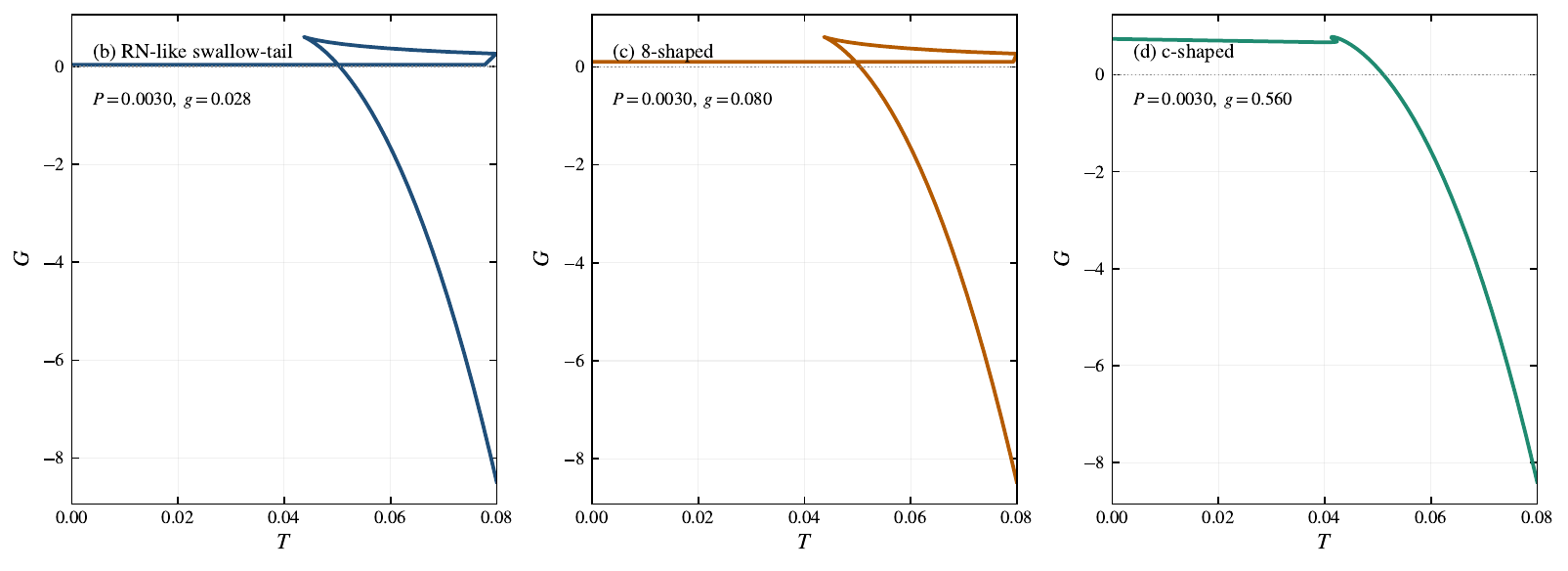}
\caption{Representative direct Bardeen-AdS Gibbs curves at $P=0.0030$. The panels display the RN-AdS-like, $8$-shaped, and c-shaped topologies obtained as $g$ increases.}
\label{fig:gt-}
\end{figure*}

\begin{figure}[t]
\centering
\includegraphics[width=0.95\linewidth]{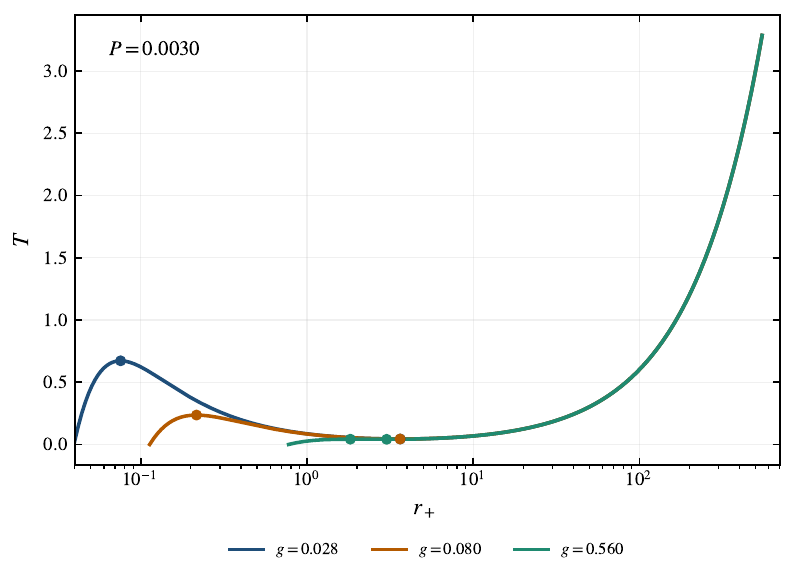}
\caption{Direct Bardeen-AdS temperature curves at $P=0.0030$ for the same representative values of $g$ as in Fig.~\ref{fig:gt-}. The markers indicate the positive turning points used in the topology classifier.}
\label{fig:tr-}
\end{figure}

At $P=0.0030$, the representative curves in Fig.~\ref{fig:gt-} display the three multibranch topologies realized along this pressure slice. For $g=0.028$, the Gibbs curve remains RN-AdS-like. Slightly above the primary threshold, at $g=0.080$, the curve becomes $8$-shaped. At $g=0.560$, the two-intersection structure has disappeared and the curve is c-shaped. The corresponding temperature curves in Fig.~\ref{fig:tr-} give the turning points underlying the classification.

The topology change is localized in the low-temperature part of the direct Gibbs curve, with the asymptotic large-radius tail playing no dominant role. Once the $8$-shaped sector is entered, the Bardeen deformation continues to reorganize the branch structure before the curve eventually becomes single-branched.

\subsection{Equilibrium closure}

\begin{figure*}[t]
\centering
\includegraphics[width=0.98\textwidth]{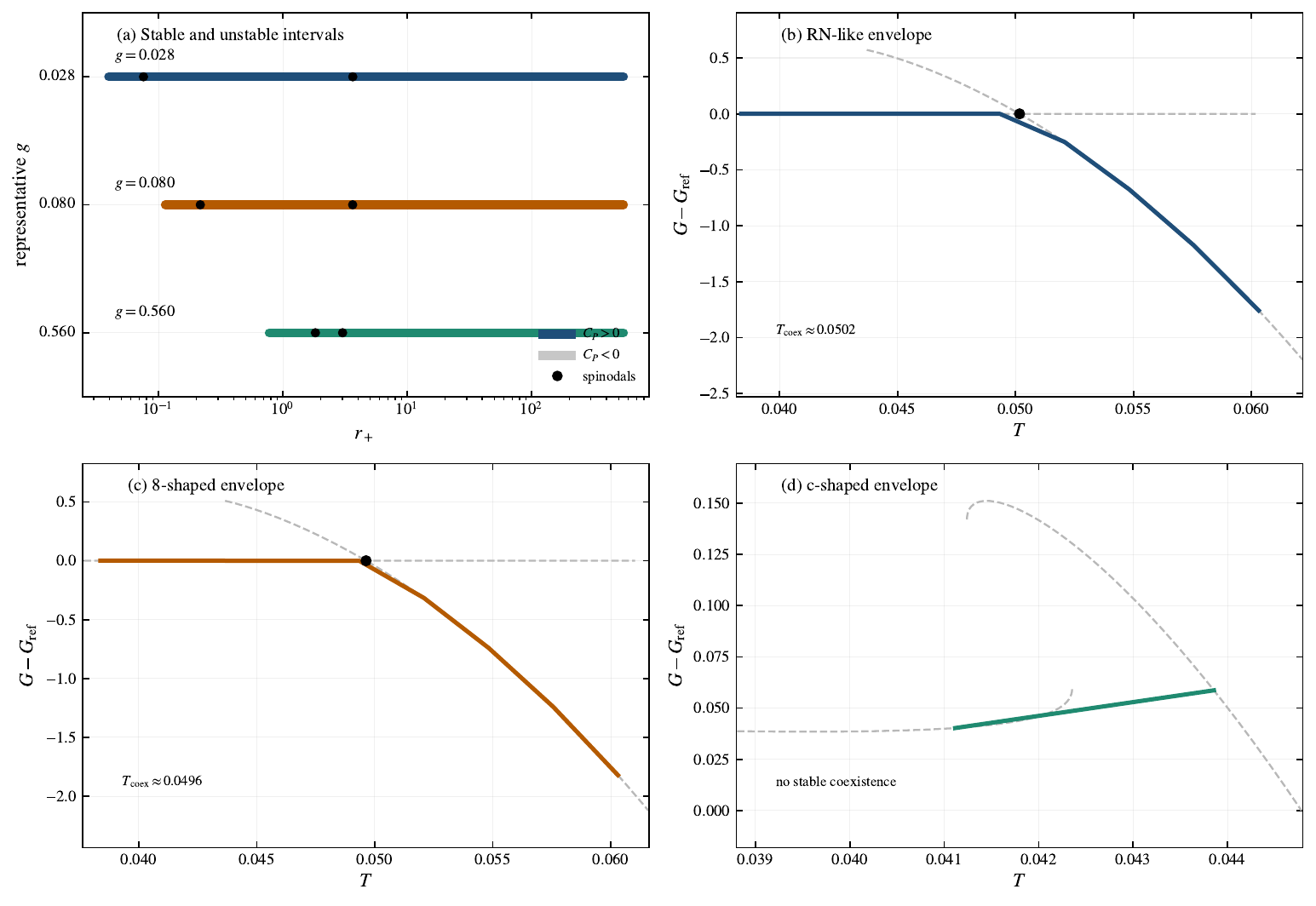}
\caption{Local stability and equilibrium closure at $P=0.0030$. The heat-capacity panel shows locally stable and unstable intervals on the positive-temperature branch, with spinodal radii marked explicitly. The stable-envelope panels show the corresponding stable-branch competition in locally shifted Gibbs coordinates for the representative RN-AdS-like, $8$-shaped, and c-shaped slices. Coexistence markers are shown where stable crossings exist.}
\label{fig:cp-}
\end{figure*}

The stability and envelope analysis in Fig.~\ref{fig:cp-} adds the local stability criterion and the lower Gibbs envelope to the raw topology classification. The heat-capacity panel shows the positive-temperature branch split into locally stable and unstable intervals, with the spinodal radii marked explicitly. In the equilibrium construction, the relevant information in $C_P$ is the sign structure and the pole locations. The remaining panels show the corresponding local stable-branch competition in shifted Gibbs coordinates for the representative RN-AdS-like, $8$-shaped, and c-shaped slices.

The raw loss of RN-AdS-like topology at $g_*(P)$ is separated from the disappearance of stable coexistence. In the direct convention, both the RN-AdS-like representative point $g=0.028$ and the immediately post-threshold $8$-shaped representative point $g=0.080$ admit a stable coexistence crossing between small and large black-hole branches. The c-shaped representative point $g=0.560$ has no stable coexistence crossing under the same stability and envelope criteria.

Table~\ref{tab:coex-} summarizes the representative coexistence data used to document the equilibrium closure. The table is a local diagnostic of the selected pressure slice; a full phase map is outside the scope of the present calculation. The first post-threshold deformation changes the raw topology while preserving stable coexistence.

\begin{table}[t]
\caption{Representative stable coexistence data at $P=0.0030$ in the direct convention.}
\label{tab:coex-}
\begin{ruledtabular}
\begin{tabular}{c c c c c}
$g$ & Topology & $T_{\rm coex}$ & $L$ & Comment \\
\colrule
0.028 & RN-AdS-like & 0.05017 & 6.13 & stable crossing present \\
0.080 & $8$-shaped & 0.04963 & 5.80 & stable crossing present \\
0.560 & c-shaped & --- & --- & stable crossing absent \\
\end{tabular}
\end{ruledtabular}
\end{table}

\subsection{Primary and secondary boundaries}

The refined threshold values at three representative pressures are
\begin{align}
g_*(0.0015) &= 0.041240, \nonumber\\
g_*(0.0025) &= 0.031944, \nonumber\\
g_*(0.0030) &= 0.029161.
\end{align}
In all three slices, the first post-threshold topology is $8$-shaped. The same pressure slices also yield the secondary boundaries
\begin{align}
g_c(0.0015) &= 0.724361, &
g_s(0.0015) &= 0.880082, \nonumber\\
g_c(0.0025) &= 0.561088, &
g_s(0.0025) &= 0.681709, \nonumber\\
g_c(0.0030) &= 0.512201, &
g_s(0.0030) &= 0.622312.
\end{align}
Thus the $8$-shaped sector occupies a finite interval in $g$, the c-shaped sector occupies a second finite interval, and only beyond $g_s(P)$ does the direct curve become single-branched.

The threshold slices are collected in Table~\ref{tab:thresholds-}. The near-constancy of the reduced combinations is visible at the table level, before the graphical collapse is shown.

\begin{table*}[t]
\caption{Exact threshold slices used in the manuscript.}
\label{tab:thresholds-}
\begin{ruledtabular}
\begin{tabular}{c c c c c c c}
$P$ & $g_*(P)$ & $g_c(P)$ & $g_s(P)$ & $\lambda_*$ & $\lambda_c$ & $\lambda_s$ \\
\colrule
0.0015 & 0.041240 & 0.724361 & 0.880082 & $6.4116\times10^{-5}$ & 0.0197807 & 0.0291996 \\
0.0025 & 0.031944 & 0.561088 & 0.681709 & $6.4116\times10^{-5}$ & 0.0197807 & 0.0291996 \\
0.0030 & 0.029161 & 0.512201 & 0.622312 & $6.4116\times10^{-5}$ & 0.0197807 & 0.0291996 \\
\end{tabular}
\end{ruledtabular}
\end{table*}

\subsection{Reduced-variable collapse}

\begin{figure*}[t]
\centering
\includegraphics[width=0.98\textwidth]{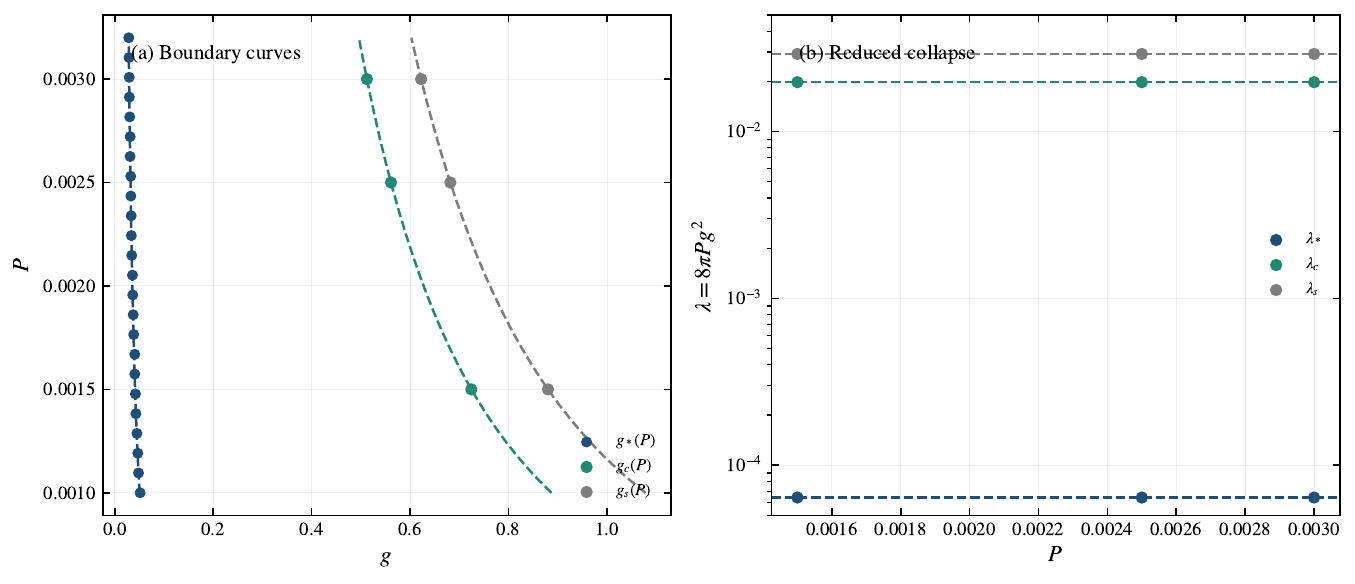}
\caption{Boundary organization in the direct fixed-pressure Bardeen-AdS scan. The boundary panel shows the primary and secondary boundaries in the $P$-$g$ plane. The reduced-variable panel shows the collapse of the corresponding combinations $\lambda=8\pi P g^2$ to pressure-independent constants.}
\label{fig:scale-}
\end{figure*}

The boundary scaling is shown in Fig.~\ref{fig:scale-}. The left panel displays the primary and secondary boundaries in the $P$-$g$ plane. The right panel shows the corresponding combinations
\begin{equation}
\lambda_* = 8\pi P g_*^2,\qquad
\lambda_c = 8\pi P g_c^2,\qquad
\lambda_s = 8\pi P g_s^2.
\end{equation}
The collapse gives
\begin{align}
\lambda_* &= 6.4116\times 10^{-5},\\
\lambda_c &= 1.97807\times 10^{-2},\\
\lambda_s &= 2.919964344\times 10^{-2},
\end{align}
where $\lambda_s$ is the exact discriminant value. Equivalently,
\begin{equation}
g_*(P)=0.001597\,P^{-1/2},
\end{equation}
\begin{align}
g_c(P)&=0.02805\,P^{-1/2},\\
g_s(P)&=0.0340854\,P^{-1/2}.
\end{align}
The direct fixed-pressure Gibbs-curve boundaries are level sets of $\lambda$ in the reduced thermodynamics. The observed $P^{-1/2}$ behavior follows from the reduced-variable formulation.

\section{Conclusion}
Within the direct Bardeen-AdS convention $G=M-TS$, the fixed-pressure scan exhibits three distinct boundary scales: the loss of the RN-AdS-like swallow-tail class at $g_*(P)$, the transition to the c-shaped sector at $g_c(P)$, and the final transition to the single-branch regime at $g_s(P)$. Adding the fixed-pressure heat-capacity criterion and the lower Gibbs envelope shows that the first of these boundaries still permits stable coexistence: the representative RN-AdS-like and immediately post-threshold $8$-shaped slices admit stable small/large crossings, while the representative c-shaped slice has no stable crossing.

After the rescaling $r_+=g\rho$, the direct temperature, direct Gibbs potential, and turning-point equation depend only on $\lambda=8\pi P g^2$. Thus the inverse-square-root law for the boundary curves in the $P$-$g$ plane follows from the scaling structure, while the nontrivial information is contained in the constants $\lambda_*$, $\lambda_c$, and $\lambda_s$ and the ordered bifurcation sequence they define. In particular, the single-branch boundary is exact,
\begin{equation}
\lambda_s = \frac{73}{48}-\frac{13\sqrt{273}}{144}.
\end{equation}

Within the \emph{direct} horizon-based Bardeen-AdS convention, RN-AdS supplies the reference black-hole topology. In this setting, previously reported non-RN Gibbs behavior is resolved into a reproducible threshold structure with local stability, stable-envelope selection, analytic reduced control, and explicit numerical boundaries.

Further work can test whether the same equilibrium sequence survives under projected-Gibbs or restricted phase-space prescriptions, formulate the singular/regular relation inside a larger unconstrained phase space, and seek closed bifurcation conditions for $\lambda_*$ and $\lambda_c$. Such analyses would determine which parts of the threshold structure are tied to the direct convention and which persist across regular AdS black-hole thermodynamics.

\begin{acknowledgments}
The author acknowledges Scientify~\cite{scientify2026}, an AI research system developed by TsingyuAI, for leading the end-to-end technical workflow of this study. Scientify organized the literature context, derived and checked the reduced thermodynamic relations, performed the numerical scans and stability-envelope classification, generated the machine-readable evidence and figures, and prepared the manuscript and reproducibility package. The human author defined the scientific scope, reviewed the calculations and text, and assumes responsibility for the submitted work.
\end{acknowledgments}

\appendix
\section{Turning-point derivation}
The direct Bardeen-AdS temperature is
\begin{equation}
T_{\rm B}(r_+;P,g)=\frac{r_+^2-2g^2+8\pi P r_+^4}{4\pi r_+(r_+^2+g^2)}.
\end{equation}
Differentiating with respect to $r_+$ and clearing the nonzero denominator gives the turning-point condition
\begin{equation}
8\pi P r_+^6 +(24\pi P g^2-1)r_+^4 +7g^2 r_+^2 +2g^4=0.
\end{equation}
Introducing $x=r_+^2$ gives
\begin{equation}
8\pi P x^3 +(24\pi P g^2-1)x^2 +7g^2 x +2g^4 =0.
\end{equation}
With the further rescaling $x=g^2 y$ and $\lambda=8\pi P g^2$, one obtains the reduced cubic
\begin{equation}
\lambda y^3 +(3\lambda-1)y^2 +7y +2 =0.
\end{equation}
This is the turning-point equation used throughout the main text.

\section{Stability and coexistence formulas}
Using $S=\pi r_+^2$ and the chain rule, the direct fixed-pressure heat capacity is
\begin{widetext}
\begin{equation}
C_P=T\left(\frac{\partial S}{\partial T}\right)_{P,g}
=\frac{2\pi r_+^{2}(r_+^{2}+g^{2})\left(8\pi P r_+^{4}+r_+^{2}-2g^{2}\right)}
{8\pi P r_+^{6}+(24\pi P g^{2}-1)r_+^{4}+7g^{2}r_+^{2}+2g^{4}}.
\end{equation}
\end{widetext}
Its poles coincide with the turning points of $T(r_+)$, so the turning-point loci are spinodals in the direct convention.

The equilibrium Gibbs envelope is defined by
\begin{equation}
G_{\rm eq}(T;P,g)=\min_i G_i(T;P,g),
\end{equation}
where the minimum is taken over stable branches with $C_P>0$. Coexistence points are determined by
\begin{equation}
T(r_s;P,g)=T(r_l;P,g),
\qquad
G(r_s;P,g)=G(r_l;P,g),
\end{equation}
and the corresponding latent heat is
\begin{equation}
L=T\,[S(r_l)-S(r_s)].
\end{equation}

For the representative pressure $P=0.0030$, the direct-convention numerics give
\begin{itemize}
\item $g=0.028$: a stable coexistence point at $T=0.05017$ with $L=6.13$,
\item $g=0.080$: a stable coexistence point at $T=0.04963$ with $L=5.80$,
\item $g=0.560$: no stable coexistence point under the same stability and envelope criteria.
\end{itemize}
These values document the equilibrium closure used in the main text.

\section{Boundary constants and numerical settings}
The primary pressure slices are
\[
P=0.0015,\qquad P=0.0025,\qquad P=0.0030.
\]
For these slices, the direct-convention thresholds are
\begin{align}
g_*(0.0015)&=0.041240,\\
g_*(0.0025)&=0.031944,\\
g_*(0.0030)&=0.029161,
\end{align}
\begin{align}
g_c(0.0015)&=0.724361,\\
g_c(0.0025)&=0.561088,\\
g_c(0.0030)&=0.512201,
\end{align}
\begin{align}
g_s(0.0015)&=0.880082,\\
g_s(0.0025)&=0.681709,\\
g_s(0.0030)&=0.622312.
\end{align}

The corresponding reduced constants are
\begin{align}
\lambda_* &= 6.4115974685\times 10^{-5},\\
\lambda_c &= 1.9780672939\times 10^{-2},
\end{align}
with variations below the last retained digit across the pressure slices. The single-branch value is fixed by
\begin{equation}
\lambda_s=\frac{73}{48}-\frac{13\sqrt{273}}{144}
=0.029199643443347467.
\end{equation}

The accompanying source bundle includes the dense topology scan, the exact threshold slices, the representative equilibrium summary, the secondary-boundary table, and the reduced boundary constants in machine-readable form.

\section{Numerical reproducibility checks}
This appendix records the reproducibility checks used for the numerical claims in the main text. The topology classifier first constructs the positive-temperature parametric curve with coordinates
\[
(T_{\rm B}(r_+;P,g),G_{\rm B}(r_+;P,g))
\]
and counts positive temperature turning points together with branch intersections in the $G$-$T$ projection. The RN-AdS-like class is defined operationally by the one-intersection swallow-tail pattern of the RN-AdS reference curve. The $8$-shaped, c-shaped, and single-branch labels are then assigned by the corresponding intersection and turning-point patterns.

The primary threshold $g_*(P)$ is obtained by refining the first value of $g$ at which the RN-AdS-like label is lost at fixed pressure. The secondary threshold $g_c(P)$ is obtained by refining the transition from the $8$-shaped sector to the c-shaped sector, and $g_s(P)$ is obtained from the merger of the positive turning points. All thresholds reported in Table~\ref{tab:thresholds-} are stored in the machine-readable tables supplied with the source package.

The reduced constants are checked by recomputing
\[
\lambda_i=8\pi P g_i(P)^2
\]
for each exact pressure slice. The analytic value of $\lambda_s$ is independently recomputed from
\[
\lambda_s=\frac{73}{48}-\frac{13\sqrt{273}}{144}.
\]
The verification script included with the source package checks these identities, confirms the coexistence values quoted in Table~\ref{tab:coex-}, and verifies that all figure files referenced by the manuscript are present.

\bibliographystyle{apsrev4-2}
\bibliography{references}

@article{hawkingpage1983,
  title = {Thermodynamics of Black Holes in Anti-de Sitter Space},
  author = {Hawking, S. W. and Page, Don N.},
  journal = {Communications in Mathematical Physics},
  volume = {87},
  pages = {577--588},
  year = {1983},
  doi = {10.1007/BF01208266}
}

@article{chamblin1999a,
  title = {Charged {A}d{S} Black Holes and Catastrophic Holography},
  author = {Chamblin, Andrew and Emparan, Roberto and Johnson, Clifford V. and Myers, Robert C.},
  journal = {Physical Review D},
  volume = {60},
  pages = {064018},
  year = {1999},
  doi = {10.1103/PhysRevD.60.064018},
  eprint = {hep-th/9902170},
  archivePrefix = {arXiv}
}

@article{chamblin1999b,
  title = {Holography, Thermodynamics and Fluctuations of Charged {A}d{S} Black Holes},
  author = {Chamblin, Andrew and Emparan, Roberto and Johnson, Clifford V. and Myers, Robert C.},
  journal = {Physical Review D},
  volume = {60},
  pages = {104026},
  year = {1999},
  doi = {10.1103/PhysRevD.60.104026},
  eprint = {hep-th/9904197},
  archivePrefix = {arXiv}
}

@article{kubiznak2012,
  title = {{P-V} criticality of charged {A}d{S} black holes},
  author = {Kubiz\v{n}\'ak, David and Mann, Robert B.},
  journal = {Journal of High Energy Physics},
  volume = {2012},
  number = {7},
  pages = {33},
  year = {2012},
  doi = {10.1007/JHEP07(2012)033},
  eprint = {1205.0559},
  archivePrefix = {arXiv}
}

@article{mann2017,
  title = {Black hole chemistry: thermodynamics with {Lambda}},
  author = {Kubiz\v{n}\'ak, David and Mann, Robert B. and Teo, Mae},
  journal = {Classical and Quantum Gravity},
  volume = {34},
  number = {6},
  pages = {063001},
  year = {2017},
  doi = {10.1088/1361-6382/aa5c69},
  eprint = {1608.06147},
  archivePrefix = {arXiv}
}

@misc{bardeen1968,
  title = {Non-singular general relativistic gravitational collapse},
  author = {Bardeen, James M.},
  year = {1968},
  note = {Proceedings of GR5, Tbilisi, USSR, p.~174}
}

@article{ayonbeato2000,
  title = {The {Bardeen} model as a nonlinear magnetic monopole},
  author = {Ay\'on-Beato, Eloy and Garc\'ia, Alberto},
  journal = {Physics Letters B},
  volume = {493},
  pages = {149--152},
  year = {2000},
  doi = {10.1016/S0370-2693(00)01125-4},
  eprint = {gr-qc/0009077},
  archivePrefix = {arXiv}
}

@article{fanwang2016,
  title = {Construction of regular black holes in general relativity},
  author = {Fan, Zhong-Ying and Wang, Xiaobao},
  journal = {Physical Review D},
  volume = {94},
  pages = {124027},
  year = {2016},
  doi = {10.1103/PhysRevD.94.124027},
  eprint = {1610.02636},
  archivePrefix = {arXiv}
}

@article{fan2017,
  title = {Critical phenomena of regular black holes in anti-de {S}itter space-time},
  author = {Fan, Zhong-Ying},
  journal = {European Physical Journal C},
  volume = {77},
  pages = {266},
  year = {2017},
  doi = {10.1140/epjc/s10052-017-4830-9},
  eprint = {1609.04489},
  archivePrefix = {arXiv}
}

@article{tzikas2019,
  title = {{Bardeen} black hole chemistry},
  author = {Tzikas, Athanasios G.},
  journal = {Physics Letters B},
  volume = {788},
  pages = {219--224},
  year = {2019},
  doi = {10.1016/j.physletb.2018.11.036},
  eprint = {1811.01104},
  archivePrefix = {arXiv}
}

@article{li2019,
  title = {Thermodynamics of the {Bardeen} black hole in anti-de {S}itter space},
  author = {Li, Cong and Fang, Chao and He, Miao and Ding, Jiacheng and Li, Ping and Deng, Jian-Bo},
  journal = {Modern Physics Letters A},
  volume = {34},
  number = {40},
  pages = {1950336},
  year = {2019},
  doi = {10.1142/S021773231950336X},
  eprint = {1812.02567},
  archivePrefix = {arXiv}
}

@article{ma2014,
  title = {Corrected form of the first law of thermodynamics for regular black holes},
  author = {Ma, Meng-Sen and Zhao, Ren},
  journal = {Classical and Quantum Gravity},
  volume = {31},
  number = {24},
  pages = {245014},
  year = {2014},
  doi = {10.1088/0264-9381/31/24/245014},
  eprint = {1411.0833},
  archivePrefix = {arXiv}
}

@article{guomiao2022,
  title = {Weinhold geometry and thermodynamics of {Bardeen} {A}d{S} black holes},
  author = {Guo, Yang and Miao, Yan-Gang},
  journal = {Nuclear Physics B},
  volume = {980},
  pages = {115839},
  year = {2022},
  doi = {10.1016/j.nuclphysb.2022.115839},
  eprint = {2107.01866},
  archivePrefix = {arXiv}
}

@article{simovic2024,
  title = {Euclidean and Hamiltonian thermodynamics for regular black holes},
  author = {Simovic, Fil and Soranidis, Ioannis},
  journal = {Physical Review D},
  volume = {109},
  pages = {044029},
  year = {2024},
  doi = {10.1103/PhysRevD.109.044029},
  eprint = {2309.09439},
  archivePrefix = {arXiv}
}

@article{ma2025,
  title = {From singular to regular: revisiting thermodynamics of {Bardeen}-{A}d{S} black holes},
  author = {Ma, Meng-Sen and He, Yun and Wang, Xiao-Ming and Li, Huai-Fan},
  journal = {Physics Letters B},
  volume = {870},
  pages = {139961},
  year = {2025},
  doi = {10.1016/j.physletb.2025.139961},
  eprint = {2510.06576},
  archivePrefix = {arXiv}
}

@article{ladghami2025,
  title = {Thermodynamics of {Bardeen} black holes in restricted phase space},
  author = {Ladghami, Yahya and Asfour, Brahim and Bouali, Amine and Errahmani, Ahmed and Ouali, Taoufik},
  journal = {Physics Letters B},
  volume = {864},
  pages = {139418},
  year = {2025},
  doi = {10.1016/j.physletb.2025.139418}
}

@article{hennigar2025,
  title = {Thermodynamics of regular black holes in anti-de {S}itter space},
  author = {Hennigar, Robie A. and Kubiz\v{n}\'ak, David and Murk, Sebastian and Soranidis, Ioannis},
  journal = {Journal of High Energy Physics},
  volume = {2025},
  number = {11},
  pages = {121},
  year = {2025},
  doi = {10.1007/JHEP11(2025)121},
  eprint = {2505.11623},
  archivePrefix = {arXiv}
}

@misc{scientify2026,
  title = {Continuous Knowledge Metabolism: Generating Scientific Hypotheses from Evolving Literature},
  author = {Tao, Jinkai and Wang, Yubo and Liu, Xiaoyu and Yang, Menglin},
  year = {2026},
  eprint = {2604.12243},
  archivePrefix = {arXiv},
  primaryClass = {cs.CL},
  doi = {10.48550/arXiv.2604.12243}
}

\end{document}